\documentclass[prl,aps,10pt,twocolumn]{revtex4-1}

\usepackage{epsfig,graphicx,amsmath}
\usepackage{amsfonts}
\usepackage{amsmath}
\usepackage{enumerate}		
\usepackage{amsfonts}
\usepackage{amssymb}
\usepackage{amsthm}

\usepackage{afterpage}
\usepackage{color}

\usepackage[caption=false]{subfig}
\newcommand{\be}{\begin{equation}}
\newcommand{\ee}{\end{equation}}
\newcommand{\bea}{\begin{eqnarray}}
\newcommand{\eea}{\end{eqnarray}}

\renewcommand{\(}{\left(}
\renewcommand{\)}{\right)}

\newcommand{\bc}{\begin{center}}
\newcommand{\ec}{\end{center}}

\newcommand{\beq}{\begin{eqnarray}}
\newcommand{\eeq}{\end{eqnarray}}
\newcommand{\beqq}{\begin{eqnarray*}}
\newcommand{\eeqq}{\end{eqnarray*}}

\newcommand{\vect}[1]{\boldsymbol{#1}}

\begin{document}
	\title{Encounter time of two chromatin loci is governed by polymer de-condensation and modification of local interactions}
	\author{A. Amitai$^{2}$ and D. Holcman$^{2}$}
	\affiliation{$^{1}$Departments of Chemical Engineering; Institute for Medical Engineering $\&$ Science, Massachusetts Institute of Technology, Cambridge 02139, MA, USA \\
	$^{2}$Institute of Biology, Ecole Normale Sup\'erieure, 46 rue d'Ulm 75005 Paris, France.}
\begin{abstract}
The time for a DNA sequence to find its homologous depends on a long random search process inside the cell nucleus. Using polymer models, we model and compute here the mean first encounter time (MFET) between two sites located on two different polymer chains and confined by potential wells. We find that reducing the potential (tethering) forces results in a local polymer decondensation near the loci and numerical simulations of the polymer model show that these changes are associated with a reduction of the MFET by several orders of magnitude. We derive here new asymptotic formula for the MFET, confirmed by Brownian simulations. We conclude that the acceleration of the search process after local chromatin decondensation can be used to analyze the local search step during homology search.
\end{abstract}
\maketitle
\begin{figure}[http!]
 \centering
 \includegraphics[width=1\linewidth]{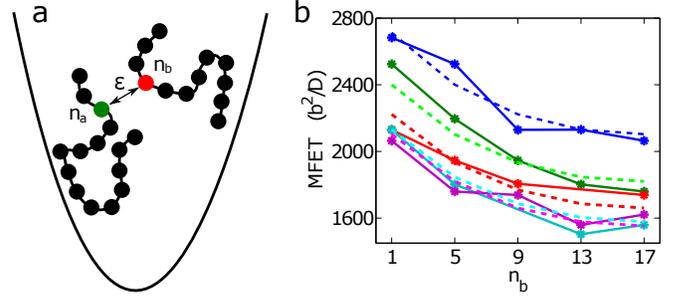}
 \caption{MFET for two monomers in a harmonic potential. ({\bf a}) Illustration of two Rouse polymers in a harmonic potential. In the model monomers $n_a$ and $n_b$ interact, when they enter the ball of radius $\epsilon$. ({\bf b}) MFET for two monomers $n_a$ and $n_b$ belonging to two different polymers (of length $N=33$) in a harmonic well of strength $\rho=0.01b^{-2}$, while $\epsilon=0.01b$. The x-axis represents the interacting monomer $n_b$, while the different curved shown for different values of $n_a$ (top to bottom): $n_a=1$ (blue), $n_a=5$ (green), $n_a=9$ (red), $n_a=13$ (cyan), $n_a=17$ (magenta). The theoretical curves (dashed lines) match the simulation results and were computed from eq.42 with $\rho=0.01b^{-2}$ and $C=190b^2/D$ was fitted.
} \label{fig:MFET_harmonic}
  \end{figure}

The repair of DNA double-strand breaks (DSBs) is key for cell survival, however the underlying physical mechanisms remains difficult to describe, mostly because they involve multiple spatial scales. One repair pathway is homologous recombination (HR), where broken strands have to perform a random search for a homologous DNA template, that will be used to repair the break \cite{Kupiec2008}. Analysis of single particle trajectories of DNA loci \cite{Bronstein2009,Bronshtein2015} before and after DSB induction revealed that chromatin decondenses by releasing tethering forces \cite{Amitai2017}, which may have consequences for the efficient search time. We present here polymer modeling and analytical tools for understanding this specific search step.\\
Search processes involving loci located on polymers have been investigated in the context of polymer looping \cite{Wilemski1973,Wilemski1974_1,Sokolov2003,Amitai2012_1}. However, much less is known about the mean time to meet for two monomers that belong to two different polymers. The difficulties of an analytical treatment \cite{Pastor1996} are due to the multiple relaxation times that cannot be reduced to a single Brownian particle diffusing in a potential well, representing the end-to-end distance energy \cite{Szabo1981}. Historically, the encounter between two monomers on different polymers \cite{deGennes1982}  was reduced to the dynamics a single particle with an effective time dependent diffusivity.  This simplification is too drastic and thus cannot be used for predicting the local chromatin reorganization, especially in confinement domains \cite{Amitai2013_1}.\\
To represent the changes in chromatin organization following DSB and study the random search dynamics, polymer models are well suited, starting with the Rouse model \cite{Doi:Book}, followed by more recent developments such as $\beta$-polymer \cite{Amitai2013_2} and more accurate ones are now used in stochastic simulations \cite{Heermann2011,Heermann2012_2,Heermann2012_1,Amitai2017}. Although a multiscale approach would be necessary to integrate both refined and coarse-grained behavior of the chromatin fiber, details about local protein organization is still missing and we are thus focusing here on the local search time for homologous sequences located on two different polymers restricted by external interactions, that we model as potential wells.  We derive asymptotic formula, that we verify with stochastic simulations. Deriving these formulas reduces the exploration of parameter space and explains the complex dependency of the search time with respect to the main physical parameters. Although this step is far from covering the entire HR mechanism, it sheds some light on the first step of HR and shows how physical constrains affect the search process.\\
{\bf \noindent Search for Rouse polymer confined in a potential well.} A Rouse polymer is a collection of monomers with positions $R=[r_1,r_2,...,r_N]^T$, connected sequentially by harmonic springs \cite{Doi:Book}. We consider here two chains of the same length $N$ where monomers are positioned at $\vect R_{i,n}$ ($n=1,2,...N$, $i=a,b$), driven by a random Brownian motion and coupled to a spring force originating from the nearest neighbors. There are no direct forces between the two chains. The potential energy of the chains is the sum
{\small
\beq\label{eq:Hamiltonian}
\phi_{\rm{2ch}}= \phi^a_{\rm{Rouse}}(\vect R_{a,1},..\vect R_{a,N}) + \phi^b_{\rm{Rouse}}(\vect R_{b,1},..\vect R_{b,N})
\eeq
}
where
{\small
\beq
\phi^a_{\rm{Rouse}}(\vect R_{a,1},..\vect R_{a,N})= \frac{\kappa}{2} \sum_{n=1}^N \left( \vect R_{a,n} - \vect R_{a,n-1}\right)^2,\\
\phi^b_{\rm{Rouse}}(\vect R_{b,1},..\vect R_{b,N})= \frac{\kappa}{2} \sum_{n=1}^N \left( \vect R_{b,n} - \vect R_{b,n-1}\right)^2,
\eeq
}
where the spring constant $\kappa=3 k_B T/ b^2$ is related to the standard-deviation $b$ of the distance between adjacent monomers \cite{Doi:Book} with $k_B$ the Boltzmann coefficient and $T$ the temperature. In units of $k_B T$, we have $\kappa = 3/b^2$ and $D = 1/\gamma$, where $\gamma$ is the friction coefficient. In the Smoluchowski's limit of the Langevin equation \cite{Schuss:Book}, the dynamics of monomer $\vect R_{i,n}$ is
\beq\label{eq:Langevin_monomer_position}
\frac{d \vect{ R_{i,n}}}{d t} = -D\nabla_{\vect{ R_{i,n}}} \phi_{\rm{Rouse}} + \sqrt{2D}\frac{d\vect{w_{i,n}}}{dt},
\eeq
for $n=1,..,N$ and $i=a,b$, and $\vect w_{i,n}$ are independent three-dimensional white noises with mean zero and variance $1$. We focus here on two monomers located on two different Rouse polymers $a$ and $b$ with the same length ($N_a=N_b=N$).\\
{\noindent \bf The MFET in a harmonic potential.}
The first result concerns the search time $\langle \tau_e \rangle$ between two monomers $n_a,n_b$ to come into a distance $\varepsilon<b$ defined by the mean of the random time
\beq\label{contact_mon}
\tau_e =\inf\{ t>0 \hbox{ such that }|\vect R_{a,n_a}(t) - \vect R_{b,n_b}(t)| \leq \varepsilon\},
\eeq
where the two chains are restricted in the same potential well of strength $\rho$ acting on all monomers, so that the energy is
\beq\label{harmonic_pot}
\phi(\vect R_{a,1},..\vect R_{a,N}) &=& \phi_{\rm{Rouse}} + \frac{\rho}{2}\sum^{N}_{n=1}( \vect R_{a,n}^2+\vect R_{b,n}^2) \nonumber\\
&=& \frac{1}{2} \sum^{N-1}_{p=0} (\kappa_p+\rho) (\vect u_{a,p}^2 + \vect u_{b,p}^2),
\eeq
where $\vect u_{a,p}$ and $\vect u_{b,p}$ are the eigenvectors of the Rouse polymers $a$ and $b$ (SI eq.7). We derived (in the SI) the asymptotic formula for the two middle monomers located on two $a$ and $b$,
{\small\beq\label{tau_mid_mid}
&&\langle \tau^{\textrm{mid},\textrm{mid}}_{\epsilon} \rangle = \frac{\pi^2 }{ D \varepsilon} \Bigg[ \frac{2}{\pi\sqrt{\kappa\rho}}\tan^{-1}\(2\sqrt{\frac{\kappa}{\rho}}\tan^{-1}\(\frac{\pi}{2N} \)\) + \nonumber\\
&&\frac{2}{N(4\kappa+\rho)}\Bigg]^{3/2}+C(\epsilon,N),
\eeq
} while for the end ones,
{\small
\beq\label{tau_end_end}
&&\langle \tau^{\textrm{end},\textrm{end}}_{\epsilon} \rangle  =  \frac{\pi^2}{D\varepsilon} \Bigg[ \frac{4}{\pi\sqrt{\kappa\rho}}\tan^{-1}\(2\sqrt{\frac{\kappa}{\rho}}\tan^{-1}\(\frac{\pi}{2N} \)\) -\frac{1}{\kappa} \Bigg]^{3/2} \nonumber \\
&&+C(\epsilon,N).
\eeq
}
Note that the constant $C(\epsilon,N)$  is an order $\mathcal O(1)$ correction.  For encounter of the any two monomers, there is no closed analytical solution (see SI eq.42). To explore the range of validity of these formulas, we use Brownian simulations of two polymers confined in a harmonic well (Fig.\ref{fig:MFET_harmonic}a). As predicted by eqs.\eqref{tau_mid_mid} and \eqref{tau_end_end}, the MFET depends on the position of the interacting monomers ($n_a$ and $n_b$) along the two chains $a$ and $b$ respectively. We compare the numerical simulations with formula SI-eq.42 and fitted the constant term $C(\varepsilon,N)$ to numerical simulation results (fig.\ref{fig:MFET_harmonic}b-c): for parameters $\varepsilon=0.01b,N=33$, we found $C=190 b^2/D$. The analytical (dashed) and the numerical (points) are compared in fig.\ref{fig:MFET_harmonic}b.  The value of $C(\epsilon,N)$ contains higher order terms in the expansion of $\lambda^{\epsilon}_{0}$ (see SI eq.15).\\
We conclude that the MFET is minimal for the two middle monomers. The middle monomers are localized in a confining domain and have a smaller standard deviation of their position compared  to other monomers \cite{Amitai2012_2}. Thus, the overlap of the probability distribution function of these middle is higher compared to the others. Finally, the MFET between two monomers from two different polymers can change by 50\% depending on their relative position.\\

{\noindent \bf After chromatin decompaction, the local search time decays by two orders.}
\begin{figure}[http!]
	\includegraphics[width=0.5\textwidth]{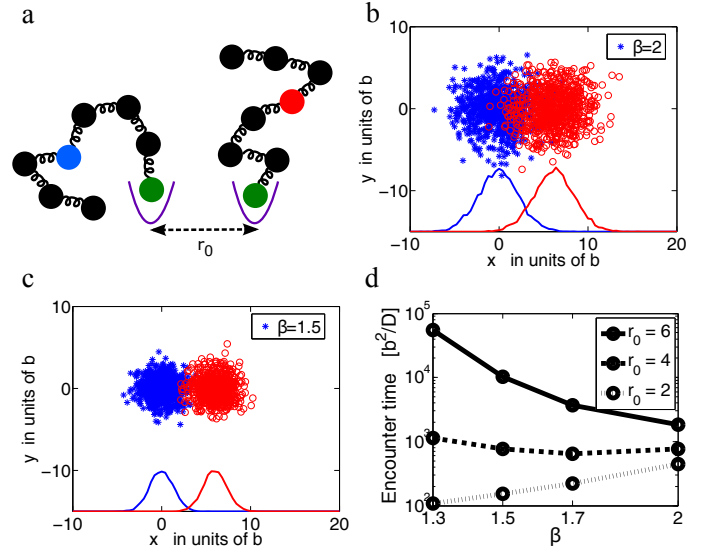}
	\caption{Condensation of a polymer chain describing chromatin. ({\bf a}) Schematic representation of two polymers with anchored extremities separated by a distance $r_0$. ({\bf b-c}) Distribution of the middle monomers belonging to different $\beta$-polymers (red asterisk and blue circles) of length $N=33$ in the XY plane. (b) $\beta=2$ (Rouse) and (c) $\beta=1.5$ ({\bf d}) MFET for the two middle monomers plotted with respect to the parameter $\beta$ and the distance $r_0$. The radius is $\epsilon=0.1b$ (unit of the time axis is $b^2/D$, where $D$ is the diffusion coefficient).}
	\label{Figure_1}
\end{figure}
Following DSB induction, the dynamics of a chromatin locus in the proximity of a break is modified, characterized by a local decondensation and thus an increase in the anomalous exponent \cite{Amitai2017}. This chromatin decondensation can be modeled by the $\beta$-model \cite{Amitai2013_2}, which describes the forces between monomers associated to a prescribed anomalous exponent in the range $]0-0.5]$. We recall that for the $\beta$-polymer model, all monomers are connected through a potential defined by
\beq
U_{\beta}(\vect R_1,..\vect R_N,\beta) = \frac{1}{2} \sum_{l,m} A_{lm} \vect R_l \vect R_m,
\eeq
with coefficients
\beq
A_{l,m} = \sum^{N-1}_{p=1} \tilde \kappa_p \alpha^l_p \alpha^m_p.
\eeq
and
\beq\label{kappa_beta}
\tilde\kappa_p = 4\kappa \sin^{\beta}\(\frac{p\pi}{2N}\) \hbox{ for } p=0..N-1.
\eeq
In such model, the strength of interaction $A_{l,m}$ between monomer $l$ and $m$ decays with the distance $|l-m|$ along the chain. By definition, $1 < \beta < 2$ \cite{Amitai2013_2} and the Rouse polymer is recovered for $\beta=2$, for which only nearest neighbors are connected. The model also relates the structure parameter $\beta$ to the anomalous anomalous exponent of a monomer by the relation $\alpha=1-1/\beta$. Thus, in a condensed polymer, the anomalous exponent of a monomer is lower than one belonging to a decondensed structure.\\
We reported recently that following break induction, the value of $\alpha$ increases from $0.3$ to $0.46$ \cite{Amitai2017}. Using the $\beta$ polymer model, we interpreted this change as a local decondensation of chromatin, confirmed in super-resolution microscopy \cite{Hauer2017,Amitai2017}. For $N=33$, the gyration radius (mean distance of the monomers to the center of mass) decreases from $\langle R_g \rangle= 2.34b$ for $\beta=2$ to $\langle R_g \rangle=1.4b$  ($\beta=1.7$) and $\langle R_g \rangle=1.21b$ for $\beta=1.5$.  We thus tested here how the polymer decondensation influences the mean first encounter time (MFET) for two middle monomers on two distinct polymers of length, when each are trapped by a potential well separated by a fixed distance $r_0$ (distance between the extremities where each polymer is anchored, fig.\ref{Figure_1}a). This scenario emulates a local homology search. We
first plotted the distribution of the middle monomers while decreasing the parameter $\beta$. We find that as $\beta$ increases, the overlap of the two distributions decreases, as shown by the steady-state distributions of each monomer (Fig.\ref{Figure_1}b,c), obtained for various value o $\beta$. \\
Next, we estimated the MFET of the middle monomers of each chain for monomers $n_a=n_b=17$. We find that increasing $\beta$ from 1.3 to 2, leading to a decondensed polymer, decreases the MFET by two orders of magnitude (Fig.\ref{Figure_1}d, thick black line). This situation corresponds to a decondensation, where tethering forces, characterized by the spring constant $\approx 0.36k_BT/b^2$ between the middle monomer and its two neighbors. Finally, changing the anchoring distance $r_0$ affects the MFET by even larger orders of magnitude (Fig.\ref{Figure_1}h).\\
Interestingly, as $\beta$ increases, the MFET does not always decrease. For small anchoring distances $r_0=2b$, decreasing $\beta$ represent a condensed polymer, where the searching monomers have more chances to meet thus leading to a reduced MFET (Fig.\ref{Figure_1}d).
We conclude that the encounter process can be regulated by the average distance between monomers.\\
Possibly, after DSB induction, characterized by an increase in $\beta$ \cite{Amitai2017}, the distribution of loci (monomers) is modified as described in fig.\ref{Figure_1}b-c, in a manner that depends on the tethering distance between polymers. In S phase and G2,  the cohesin molecule maintains the sister chromatids together \cite{Revenkova2006}. Due to their close proximity, sister chromatids can thus be used as template for repair \cite{Zickler2015}. In general, the distribution of distances between two homologous loci is difficult to estimate. The median distance between cohesin binding sites is of the order of 16.5kb for chromosome II of the budding yeast \cite{Schmidt2009}. A length $b=30$nm represents a monomer of size 3kb \cite{Amitai2013_1}, and 16.5kb is a chain of about 6 monomers. The present result suggests that chromatin decondensation can significantly facilitate the encounter between monomers and we predicted here a decay of the search by 2 orders of magnitude.\\
\begin{figure}[http!]
	\centering
	\includegraphics[width=1.0\linewidth]{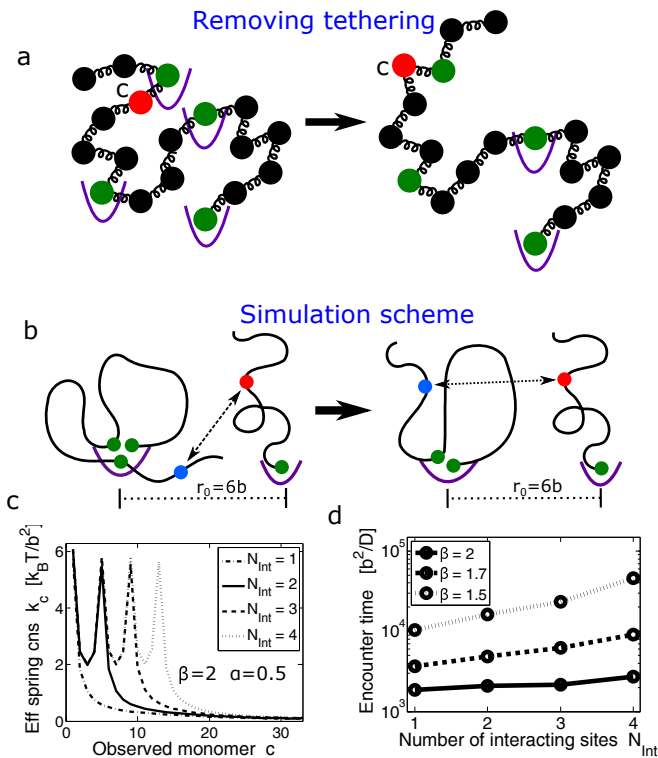}
	\caption{Loss of local tethering interactions alters the MFET. ({\bf a}) Scheme of two homologous polymers, where a fraction of monomers (green) are interacting, modeled by a fixed harmonic potential wells (purple). Following a DSB, these interactions are removed. ({\bf b}) Scenario with two polymers (length $N=33$), where the two middle monomers (blue and red) meet: only one extremity ($n=1$) belonging to one polymer is interacting with a potential well, for the other one, several monomers (green) are trapped ($n=5,9,13$) at a well located at a fixed distance. The total number of interacting monomer is $N_{Int}$. ({\bf c}) Estimation of the effective spring $k_c$ for different monomers along the chain ($c=1..33$) using formula SI-43 for a Rouse polymer. One extremity interacts with a well (dashed-pointed line). We increase the number of interacting monomers $N_{Int}$, located at $n=5$ (full line), $n=9$ (dashed line) and $n=13$  (dotted line). ({\bf d}) MFET between the middle monomers for two $\beta$-polymers and an increasing number of interacting sites.}
	\label{fig:kc}
\end{figure}
{\bf \noindent Removing tethering forces on a polymer facilitates the local search.}
Chromatin strands are well localized \cite{Zimmer2011} in the nucleus due to local interactions \cite{Amitai2015} imposed by the binding molecules such as Lamin A \cite{Bronshtein2015}. In this last section, our goal is to study the influence of local tethering forces on the search time. External forces acting on a tagged locus are characterized by a resulting tethering force with an effective spring constant $k_c$ (Fig.\ref{fig:kc}a). The consequence of this resulting force is to confine the locus motion \cite{Amitai2015}. \\
We simulated the motion of several monomers, restricted by potential wells (Fig.\ref{fig:kc}b), where the interaction energy for an interacting monomer $i$ is
\beq\label{ExternalPot}
U_i(\vect R_i)= \frac{1}{2}k (\vect R_i-\vect \mu_i)^2,
\eeq
where $\mu_i$ represents the position of the interaction and $k$ the strength of the tethering interaction. The total energy of an interacting Rouse polymer is $U_t = \phi_{\rm{2ch}} + \sum_{i\; interacting }U_i$, where $\phi_{\rm{2ch}}$ is defined in eq.\eqref{eq:Hamiltonian}. The resulting force on a tracking locus (red bead in Fig.\ref{fig:kc}a) is computed \cite{Amitai2015} by 
\beq\label{EpotentialWell2}
\lim_{\Delta t \rightarrow 0 } \mathbb E \{\frac{\vect R_c(t+\Delta t)-\vect R_c(t)}{\Delta t} | \vect R_c(t) = \vect x \}=& -D k_{cn} {\vect x},\nonumber &\\
\eeq
where $\mathbb E\{.| \vect R_c = \vect x  \}$ denotes averaging over all polymer configuration under the condition that the tagged monomer is at position $\vect R_c= \vect x$. When only one monomer $n$ is interacting with a potential (eq.\ref{ExternalPot}), $k_{cn}= \frac{k \kappa }{\kappa +(c-n)k}$.\\
To investigate the consequences of removing punctual interactions, we simulated the search of two monomers belonging to two different polymers each of length $N=33$. The distance between the first monomer of each polymer is $r_0=6b$. In these simulations, we allow an increasing number of monomers along one of the polymers to interact at the origin (Fig.\ref{fig:kc}b), thus restricting the motion of the observed locus. We then estimated the parameter $k_c$ from the simulated trajectories (formula SI-43) (Fig.\ref{fig:kc}c) for different $\beta$-polymers. When four monomers (at positions $n=1,5,9,13$) interact, we found (Fig.\ref{fig:kc}c) the following changes $k^{N_{\rm{Int}}=4}_{c=15}=1.08 \mathrm{k_BT}/b^2$  for $\beta=2$ (resp. $k^{N_{\rm{Int}}=4}_{c=15}=1.73$ for $\beta=1.5$). If only two monomers interact at positions $n=1,5$, the overall resulting interaction decays, characterized by $k^{N_{\rm{Int}}=2}_{c=15}= 0.29 \mathrm{k_BT}/b^2$ (resp. $k^{N_{\rm{Int}}=4}_{c=15}= 0.72 \mathrm{k_BT}/b^2$) for $\beta=2$ (resp. $\beta=1.5$).\\
To evaluate the influence of the number $N_{\rm{Int}}$ of interacting monomers
on the MFET, we run Brownian simulations to compute the MFET between two middle monomers ($n_a=n_b=17$) belonging to two different polymers of length $N=33$. We find that the MFET increases with $N_{\rm{Int}}$ when the two polymer extremities are separated by a distance $r_0$ (Fig.\ref{fig:kc}b). Interestingly, this increase depends on the polymer compaction, characterized by the parameter $\beta$: when $\beta=1.5$, an increase of interacting monomers from $N_{\rm{Int}}=1$ to $N_{\rm{Int}}=4$ resulted in a 4.4 fold increase of the MFET (Fig.\ref{fig:kc}d). We conclude that decreasing the tethering constant $k_c$, which represents the number of interactions of the DNA around the DSB (characterized by $N_{\rm{Int}}$), results in a drastic reduction of the encounter time between two searching homologous sites.\\
Following DSB induction, experimental data confirmed that $k_c$ is significantly decreased \cite{Amitai2017}. Thus, changes in the anchoring forces modify the locus dynamics that reflects the chromatin organization \cite{Amitai2015} (Fig.\ref{fig:kc}a).\\
To conclude, we derived here MFET formulas \ref{tau_mid_mid} and \ref{tau_end_end} for the search time between two monomers. The MFET depends on the position of the monomers, chromatin condensation (parameter $\beta$ and the anomalous exponent $\alpha$) and tethering forces that represent binding forces, mediated by molecules such as CTCF or cohesin or other protein-protein interactions. The combination of polymer decondensation and releasing tethering forces can modulate and accelerate drastically the search time in a local environment (modeled here by a potential well). Polymer decondensation (increase in $\beta$) leads to a reduced MFET. Loss of  local connectivities between monomers may arise from histone acetylation and local nucleosome eviction, as reported for DSBs in yeast \cite{Renkawitz2014}. This is likely to change the chromatin condensation and could accelerate the local search for homology. Previous chromosome capture data in yeast, performed before and after break induction \cite{Oza2009}, shows that  inter-and intra-chromosomal sites at DSB are less frequent compared to an uncleaved locus (see Fig.1a of \cite{Oza2009}).

\bibliographystyle{apsrev4-1}
\bibliography{RMPbiblio5}

\end{document}